\documentclass[11pt,twoside]{article}

\usepackage{asp2006}
\usepackage{epsf}
\usepackage{psfig}
\usepackage{lscape}
\usepackage{graphicx}

\begin{document}
\title{The importance of the remnant's mass for VLTP born again
  times. Implications for V4334 Sgr and V605 Aql.}   
\author{Miller Bertolami M. M.$^{1,2}$ and Althaus L. G.$^{1,2}$}   
\affil{$^1$ Facultad de Ciencias Astron\'omicas y Geof\'{\i}sicas, UNLP,
  Argentina\\ $^2$ Instituto de Astrof\'{\i}sica La Plata, UNLP-CONICET, Argentina}    

\begin{abstract} 
We present numerical simulations of the very late thermal pulse (VLTP)
scenario for a wide range of remnant masses. We show that, by taking into
account the different possible remnant masses, the fast outburst evolution of
V4334 Sgr (a.k.a. Sakurai's Object) and V605 Aql can be reproduced within
standard 1D stellar evolution models. A dichotomy in the born again timescales
is found, with lower mass remnants evolving in a few years and higher mass
remnants (M $\hbox{\rlap{\hbox{\lower4pt\hbox{$\sim$}}}\hbox{$>$}}$ 0.6
M$_{\odot}$) failing to expand due to the H-flash and, as a consequence,
evolving in timescales typical of He-shell flash driven born agains ($\sim
100$ yr).


\end{abstract}

\section{Introduction and description of the present work}
Since the discovery of post-AGB thermal pulses in the early simulations of
Paczynski (1970) these events have been object of several studies. Fujimoto
(1977) showed that during thermal pulses penetration of the pulse driven
convective zone (PDCZ) into the H-rich envelope is possible if the mass of
this envelope was small enough. These works were confirmed by Sch\"onberner
(1979) who performed calculations of low mass stars with steady mass
loss. Sch\"onberner (1979) found that
contact between the PDCZ and the H-rich envelope was possible if the thermal
pulse happens when the star enters the white dwarf stage (event now termed
VLTP).  Since the work of Renzini (1979) post-AGB thermal pulses are
associated with the formation of H-deficient stars, an idea pushed forward by
the simulations of Iben et al. (1983) (who introduced the term ``born again
AGB stars'' for this objects). Due to numerical difficulties in the treatment
of the simultaneous mixing and burning of H at high temperatures during a
VLTP, early simulations did not follow the evolution during this event. This
was first attempted by Iben \& MacDonald (1995) who presented the first VLTP
simulations that included a time dependent treatment of the convective mixing
and burning of H. After the works of Herwig et al. (1997, 1999) it became
clear that the abundances of some post-AGB H-deficent stars could be well
understood within the born again scenario (see, however, De Marco this
proceedings).
\begin{table*}[ht!]
{\footnotesize
\begin{center}
\begin{tabular}{c|c|c|c|c|c}
         & Iben \&   &  Herwig   &  Herwig  & Lawlor \&  & Miller Bertolami \\             
         & MacDonald &  et al.   &          & MacDonald  &  et al.    \\
         &  (1995)   &  (1999)   &  (2001)  &  (2002)    & (2006)     \\ \hline      
Mass  [M$_{\odot}$] & 0.60  & 0.604  & 0.535 & 0.56 to 0.61 &  0.589   \\ 
Born Again Time [yr] & 17 & 350 & 21 & 4.5 to 8.5 & 5 to 10 \\ 
\end{tabular}
\label{tabla}
\caption{Born Again timescales of previous simulations of the VLTP with
  standard MLT approach that include the violent H-burning event.}
\end{center}
}
\end{table*}

The identification of V4334 Sgr as a star undergoing a VLTP event (Duerbeck \&
Benetti 1996) has renewed the interest in this particular kind of late helium
shell flash. V4344 Sgr has showed a very fast evolution of a few years
(Duerbeck et al. 1997, Asplund et al. 1999 and van Hoof et al. 2007), not far
from the only theoretical value available at the time of its discovery ($\sim
17$ yr, Iben \& MacDonald 1995). Since then, the fast evolution of V4334 Sgr
has been the object of several works with results far from showing a
consistent picture (see Table 1). The works of Herwig (2001) and Hajduk et
al. (2005) introduced a new free parameter in the treatment of convection (the
mixing efficiency) and showed that, by fine tuning the velocities of
convective mixing during the VLTP (to values different to those predicted by
the mixing length theory, MLT), the fast outburst evolution and reheating of
V4334 Sgr could be reproduced by a model of 0.604 M$_{\odot}$. In this context,
VLTP evolution was proposed as a tool to test convection theory of reactive
convective fluids (Herwig 2001). However, born again timescales of previous
simulations of the VLTP episode with standard MLT approach (i.e. no reduction
in convective velocities) show a wide range of born again timescales, with
differences rising up to a factor 70 (see Table 1). If V4334 Sgr observed
evolution and VLTP simulations are to be used to test convection theory, these
differences have to be understood first.

As can be seen from third and fourth columns of Table 1 the remnant's mass
seems to play an important role in determining the born again
timescale. However, most of the existing simulations of the VLTP have been
performed in a narrow range in mass. The aim of the present article is the
exploration of the (neglected) importance of the remnant mass for the born
again timescale.


For the present work we have performed numerical simulations of the VLTP for
10 different remnant masses. All of the remnant models are the result of full
 evolutionary calculations from the ZAMS, through the AGB phase
and to the post-AGB phase for a metallicity of Z=0.02.  Overshooting at every
convective boundary was considered as in Herwig et al. (1997). Convective
mixing was considered as a difussive process and solved simultaneously with
nuclear burning and convective
velocities were adopted as given by standard MLT approach.
For a more extensive discussion about numerical and physical aspects of
the present simulations we refer the reader to Miller
Bertolami \& Althaus (2007).

\section{Discussion of the results and comparison with observations}
As can be seen in Fig. 1 (right panel) our sequences show a clear dichotomy in
the VLTP born again timescales as a function of mass. While remnants with
masses $M\hbox{\rlap{\hbox{\lower4pt\hbox{$\sim$}}}\hbox{$<$}}0.6$M$_{\odot}$
show a fast born again evolution of a few years, higher mass remnants display
born again timescales of the order of centuries, timescales typical of
He-flash driven born again evolutions --- i.e. when no burning of H takes
place (see Bl\"ocker 1995 and Sch\"onberner, this proceedings). The reason for
this can be found in the estimation presented in Fig. 1., where the energy
available from the burning of the whole H-content of the star ($E_{\rm H}$) is
compared with the energy needed to expand the envelope above the point at
which such energy is released ($E_{\rm exp}$). As it is shown there, for
remnants above a certain value ($\sim 0.6$M$_{\odot}$, thick grey line) the
energy released by the burning of the H-content of the star is not enough to
drive the expansion of the envelope. As a consequence, after burning the
H-rich envelope, the expansion back to the AGB of those sequences is driven by
the He-shell flash, leading to born again timescales of the order of
centuries.


In Fig. 2, left panel, the evolution of our low mass sequences
(M $\hbox{\rlap{\hbox{\lower4pt\hbox{$\sim$}}}\hbox{$<$}}$ 0.6 M$_{\odot}$) is
compared with observations of V4334 Sgr. As can be seen, all of them show good
agreement with the evolution of the effective temperature of V4334 Sgr. Also
the preoutburst location is compared with that inferred in V4334 Sgr (right
panel). In particular it is worth mentioning that our $0.561$M$_{\odot}$
sequence, which nicely reproduces the 1995-1998 evolution of V4334 Sgr also
reproduces the outburst lightcurve and preoutburst location for the
\emph{same} distance of $\sim 3-4$ Kpc, consistent with
independent distance determinations which place V4334 Sgr below 4.5 Kpc
(Kimeswenger 2002). However note that this model is unable to reproduce the
rapid reheating reported by van Hoof et al. (2007). Although reheating times
can be, in principle, reproduced by increasing the mass loss once the star is
back on the AGB (see Fig. 2 left bottom panel) it is also possible that the
failure of our models to reproduce the rapid reheating of V4334 Sgr is due to
the fact that hydrostatic equilibrium is explicitly broken in the outer layers
once the model is back on the AGB.

\section{Conclusion}
We have shown that post-VLTP born again times can be divided into two groups:
low mass remnants that evolve back to the AGB in a few years and high mass
remnants in which the energy released by H-burning fails to drive the expansion
back to the AGB and, consequently, show born again timescales of the order of
a century. We also showed that V4334 Sgr evolution can be understood within
the standard MLT as the VLTP evolution of a $\sim 0.56$ M$_{\odot}$
remnant. However our models fail to reproduce the rapid reheating reported by
van Hoof et al. (2007) something that may be pointing to the need of a better
treatment of the outer layers of the envelope in our models.



\begin{figure}[ht!]
\begin{center}
\includegraphics[ height=120 pt] {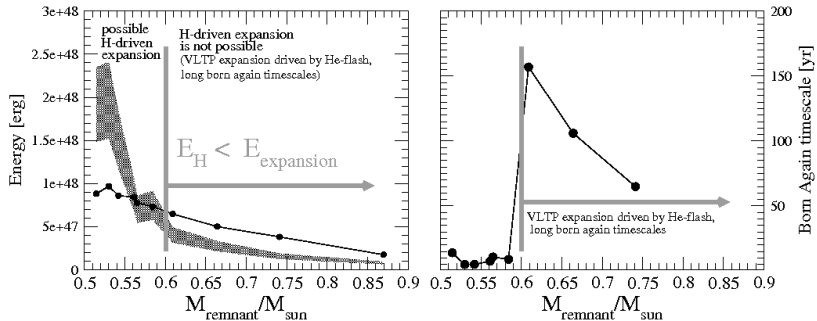}

\caption{\footnotesize Left: Comparison between the energy available from the burning of the
H-content of the star (shaded zone) with the energy needed to expand the
envelope (solid line). Right: Born again timescales of the simulations.}
\end{center}
\end{figure}

\begin{figure}[h!]
\begin{center}
\includegraphics[ height=152 pt] {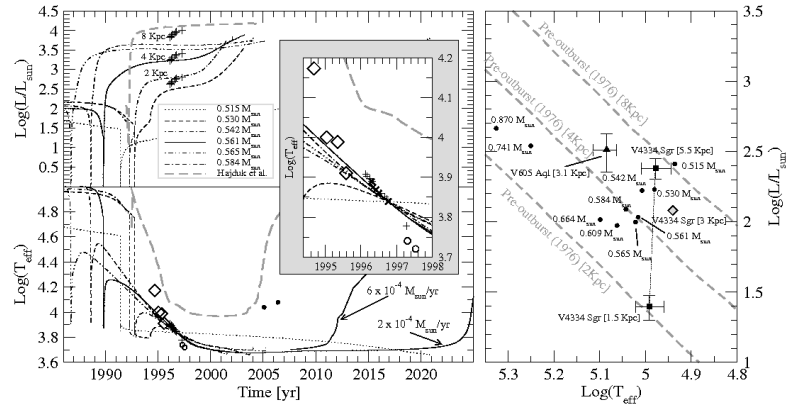}
\caption{\footnotesize \emph{Left:} Outburst evolution (top: luminosity, bottom: effective
  temperature) of our lower mass sequences as compared with observations of
  V4334 Sgr. Observations are taken from Duerbeck et al. (1997; $\diamondsuit$
  and +), Asplund et al. (1999; $\times$), Pavlenko \& Geballe (2002; $\circ$)
  and van Hoof et al. (2007; $\bullet$). For the 0.561M$_{\odot}$ model two
  different mass loss rates were considered at log($T_{\rm eff}$)$<3.8$
  leading to different reheating timescales. Inset shows a zoom of the
  evolution of the effective temperature during the discovery of V4334
  Sgr. \emph{Right:} Preoutburst location of our sequences ($\bullet$). Grey
  dashed line correspond to possible detection of V4334 Sgr in a ESO/SERC
  survey (Herwig 2001). V4334 Sgr preoutburst locations are from Kerber et
  al. (1999; black square) and Hajduk et al. (this proceedings; grey
  diamond). V605 Aql value is taken from Lechner \& Kimeswenger (2004).}
\end{center}
\end{figure}

{\footnotesize \acknowledgements M3B thanks the organizers of the Third
Conference in Hydrogen Deficient Stars for financial assistance that allowed
him to participate in the conference. We acknowledge the Monthly Notices of
the Royal Astronomical Society and Blackwell Publishing for the figures
originally published in Miller Bertolami \& Althaus (2007).  }



\end{document}